  \documentclass{gretsi}
  \usepackage[english,french]{babel}

  \usepackage{times}			
  
\usepackage{amsmath}
\usepackage{amssymb}
\usepackage{graphicx}
\usepackage{float}
\usepackage{subfigure}
\usepackage{array}
\usepackage{tabulary}
\usepackage[T1]{fontenc}

\titre{\textit{Sliced Inverse Regression} pour la d\'etermination des param\`etres stellaires fondamentaux}

\auteur{\coord{Victor}{Watson}{1,2},
        \coord{Jean-Fran\c{c}ois}{Trouilhet}{1,2},
    \coord{Fr\'ed\'eric}{Paletou}{1,2}
    \coord{Marwan}{Gebran}{3}}
\adresse{\affil{1}{Universit\'e de Toulouse, UPS-Observatoire Midi-Pyr\'en\'ees, Irap, Toulouse, France}
         \affil{2}{CNRS,  Institut de Recherche en Astrophysique et Plan\'etologie, 14 av. E. Belin, 31400 Toulouse, France}
         \affil{3}{Department of Physics and Astronomy, Notre Dame University-Louaize, PO Box 72, Zouk Mika\"el, Lebanon}}

\email{victor.watson@irap.omp.eu,
jean-francois.trouilhet@irap.omp.eu,
frederic.paletou@irap.omp.eu, mgebran@ndu.edu.lb \vspace{-.4cm}}

\resumefrancais{Nous voulons d\'{e}terminer la valeur d'un param\`{e}tre \`a partir d'un grand vecteur de donn\'{e}es associ\'e, sans conna\^{i}tre la fonction permettant de passer d'un espace \`a l'autre. La m\'{e}thode \textit{Sliced Inverse Regression} (SIR) permet de projeter un vecteur de donn\'{e}es vers un sous-espace qui est coh\'{e}rent avec la variation du param\`{e}tre \'etudi\'e. Nous proposerons, sur la base du sous-espace obtenu gr\^{a}ce \`a SIR, une m\'{e}thode pour estimer de la mani\`{e}re la plus pr\'ecise possible la valeur que prend le param\`{e}tre auquel on s'int\'eresse. }

\resumeanglais{We aim at finding the value of an explanatory variable, through its expression in a large data vector, without knowing the link function between the explanatory variable and the data-space. Sliced Inverse Regression (SIR) method, allows the projection of a data-vector onto a subspace consistent with the explanatory variable variation. We suggest a method based on the SIR subspace, that gives the most efficient estimation of an unknown explanatory variable.
\vspace{-.5cm}}

\begin{document}

\maketitle

\section{Introduction}

\vspace{-.4cm}

L'\'evolution des d\'etecteurs et de nos
capacit\'es de stockage de l'information font que la collecte des
donn\'ees astrophysiques est toujours plus massive. Des projets comme
Gaia \cite{gaia2} ou encore le LSST \cite{lsst} en sont le parfait
exemple.
Les bases de donn\'{e}es ainsi constitu\'ees seront alors utilis\'ees
pour extraire le maximum d'informations quant \`a la caract\'erisation des
objets observ\'es.
Ici, nous nous int\'eresserons \`a la question de la d\'etermination des
param\`etres stellaires fondamentaux : temp\'{e}rature effective,
gravit\'{e} de surface et m\'etallicit\'{e} (ou composition chimique
globale) d'une \'{e}toile \`{a} partir de la mesure de son spectre
\'electromagn\'etique dans le visible.
Cette d\'{e}termination doit alors \^{e}tre faite sans connaissance
d'un lien fonctionnel entre param\`{e}tres et donn\'{e}es, mais en
ayant \`a disposition soit une base de donn\'{e}es de spectres d\'ej\`a
caract\'eris\'es, soit des spectres synth\'etiques.
Il devient donc int\'{e}ressant de mettre en \oe{}uvre des
m\'{e}thodes statistiques qui, \`{a} partir de telles bases de
donn\'{e}es, peuvent automatiquement d\'{e}terminer les quantit\'{e}s
physiques d\'esir\'ees sans que la relation entre ceux-ci et les
donn\'{e}es ne soit connue. SIR \cite{li91}, avec une application encore tr\`es confidentielle dans un contexte astrophysique \cite{bernard-michel}, 
permet d'obtenir une repr\'{e}sentation des donn\'{e}es coh\'{e}rente avec le param\`{e}tre
sans connaissance de la fonction qui les lie. Nous chercherons donc la mani\`ere la plus
efficace d'utiliser les composantes du sous-espace (que nous appellerons "directions") obtenues gr\^{a}ce \`{a} SIR.

\vspace{-.6cm}

\section{Sliced Inverse Regression}

\vspace{-.3cm}

SIR recherche le sous-espace de repr\'{e}sentation d'un vecteur de donn\'{e}es $x$ qui explique au mieux la valeur d'un param\`{e}tre $y$ associ\'e. Pour construire ce sous-espace \`{a} partir d'une base de donn\'{e}es d'objets d\'ej\`a caract\'eris\'es, on commencera par diviser l'espace des donn\'{e}es en $H$ tranches (sans recouvrement et contenant chacune le m\^eme nombre d'\'echantillons) suivant l'\'{e}volution du param\`{e}tre $y$ auquel on s'int\'{e}resse. Ainsi les \'{e}chantillons de la base de donn\'{e}es sont regroup\'{e}s en tranches dans lesquelles la valeur du param\`{e}tre $y$ varie peu. Appliquer SIR permet de construire le sous-espace qui maximise la variance entre les tranches tout en gardant une variance globale r\'{e}duite. De fait, on peut y voir un parall\`{e}le avec le fonctionnement de l'analyse factorielle discriminante \cite{LDA} dans un contexte qui aurait \'{e}t\'{e} un contexte de classification.

\vspace{-.2cm}

\subsection{Principe}\label{sec_21}

\vspace{-.2cm}

Soit $\textbf{X}$ la matrice des donn\'{e}es dont chaque ligne est un \'{e}chantillon associ\'{e} \`{a} une valeur du param\`{e}tre $y$. Nous appellerons $\textbf{X}_h$ la matrice des \'{e}l\'{e}ments de $\textbf{X}$ qui appartiennent \`a la tranche $h$. Pour chaque tranche nous calculerons la moyenne $\overline{m}_h$ des lignes de $\textbf{X}_h$. 

La matrice $\textbf{X}_H$ est la matrice des barycentres des tranches, dont chaque ligne est un des vecteur $\overline{m}_h$.

$\boldsymbol{\Gamma}$ la matrice de variance-covariance inter-tranches :
\begin{equation}\label{eq_gamma}
\boldsymbol{\Gamma} = \frac{1}{H}(\textbf{X}_H - \overline{\textbf{X}}_H)^T(\textbf{X}_H -  \overline{\textbf{X}}_H),
\end{equation} 

et $\boldsymbol{\Sigma}$ la matrice de variance-covariance totale :
\begin{equation}\label{eq_gamma}
\boldsymbol{\Sigma} = \frac{1}{N}(\textbf{X} - \overline{\textbf{X}})^T(\textbf{X} -  \overline{\textbf{X}}),
\end{equation}
o\`{u} $N$ est le nombre d'\'{e}chantillons de la base de donn\'{e}e servant \`{a} construire le sous-espace, $\overline{\textbf{X}}$ est la moyenne des lignes de $\textbf{X}$ et $\overline{\textbf{X}}_H$ est la moyenne des lignes de $\textbf{X}_H$.
Les directions d\'{e}crivant le sous-espace que l'on recherche sont les solutions de :
\begin{equation}
(\boldsymbol{\Sigma}^{-1}\boldsymbol{\Gamma} - \lambda \textbf{I}_M)u = 0.
\label{eq_2}
\end{equation}  

Les solutions  de l'\'equation \ref{eq_2} correspondent aux vecteurs propres $u$ associ\'{e}s aux plus grandes valeurs propres $\lambda$ de $\boldsymbol{\Sigma}^{-1}\boldsymbol{\Gamma}$. $M$ est la dimension de l'espace de d\'epart.

Un point critique dans la mise en \oe{}uvre de SIR, est l'inversion de la matrice $\boldsymbol{\Sigma}$. La structure des bases de donn\'{e}es, fait que la matrice $\boldsymbol{\Sigma}$ n'est pas toujours de rang plein, elle est donc mal conditionn\'{e}e. Lorsque c'est le cas, nous appliquerons, sur les donn\'ees, une analyse en composante principale (PCA) \cite{Jolliffe} comme pr\'{e}-traitement \`{a} SIR pour am\'eliorer le conditionnement de $\boldsymbol{\Sigma}$ en diminuant la dimension de l'espace.

\subsection{Exemple}

Dans cet exemple l'espace de d\'epart est de dimension 4, et nous chercherons \`a trouver le sous-espace de dimension 1 liant $x$ \`a $y$ sachant que $y(x) = 2x_1x_2+x_3+0  x_4 + \epsilon$ et que les \'elements de $x$ suivent une distribution uniforme $\mathcal{U} [0,1]$. Le bruit $\epsilon \sim \mathcal{N} (0,0.01 I_4)$

\begin{figure}[H]
\centering
\includegraphics[height = 3.5cm,width =8cm]{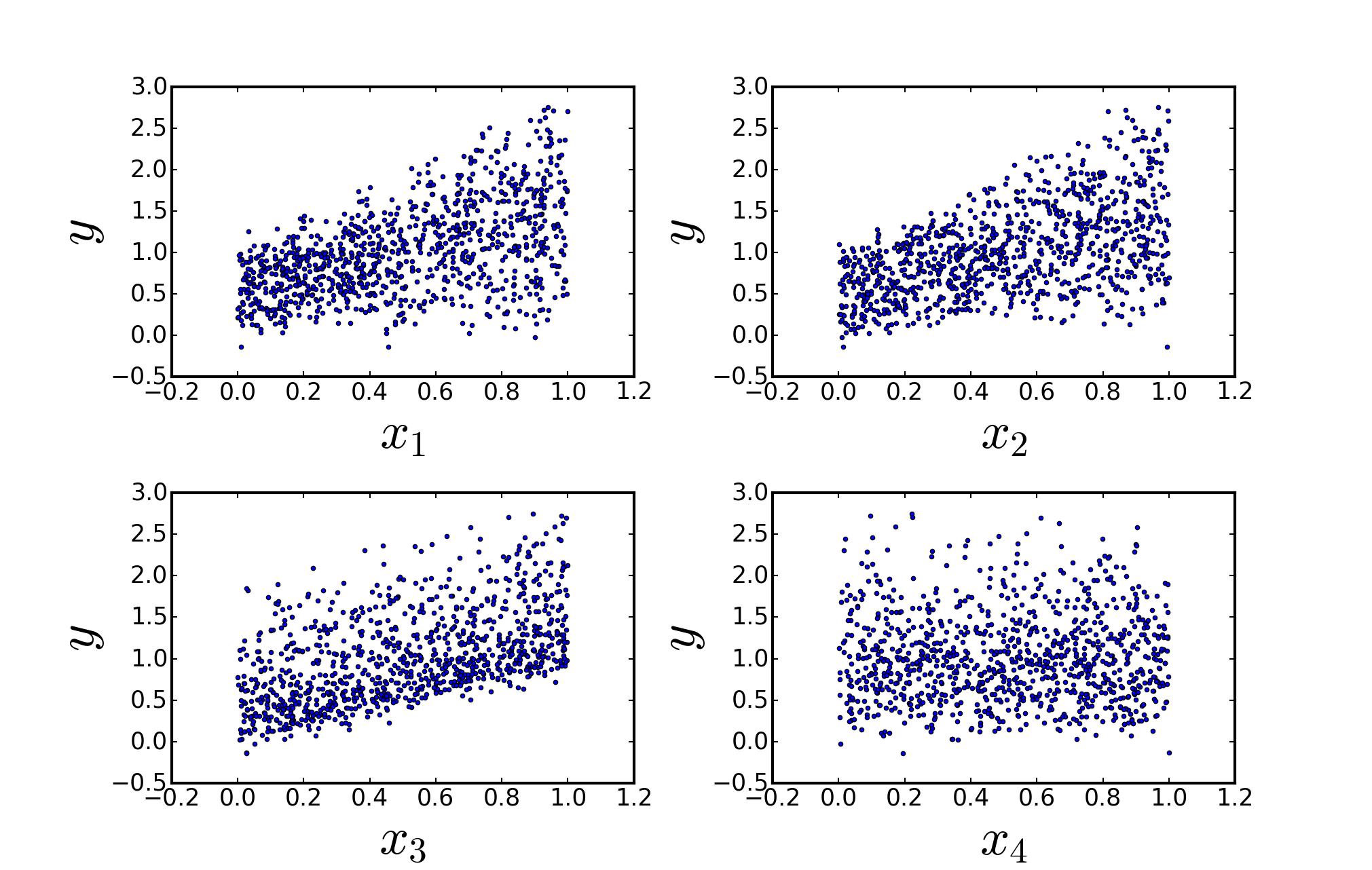}
\caption{Valeurs prises par $y$ (ordonn\'ees) par rapport aux r\'ealisations de $\textbf{X}$ (abscisses). Il n'est pas \'evident de trouver un lien entre les r\'ealisations de $\textbf{X}$ et la valeur prise par $y$ \`a partir de cet espace.}
\label{ex2}
\end{figure}

La figure \ref{ex2} montre, qu'en l'absence de la connaissance de la fonction liant $x$ \`a $y$, il est n\'ecessaire dans cet exemple  de trouver une direction permettant de retrouver la valeur de $y$.

%

\begin{figure}[H]
\centering
\includegraphics[height = 3.5cm,width =8cm]{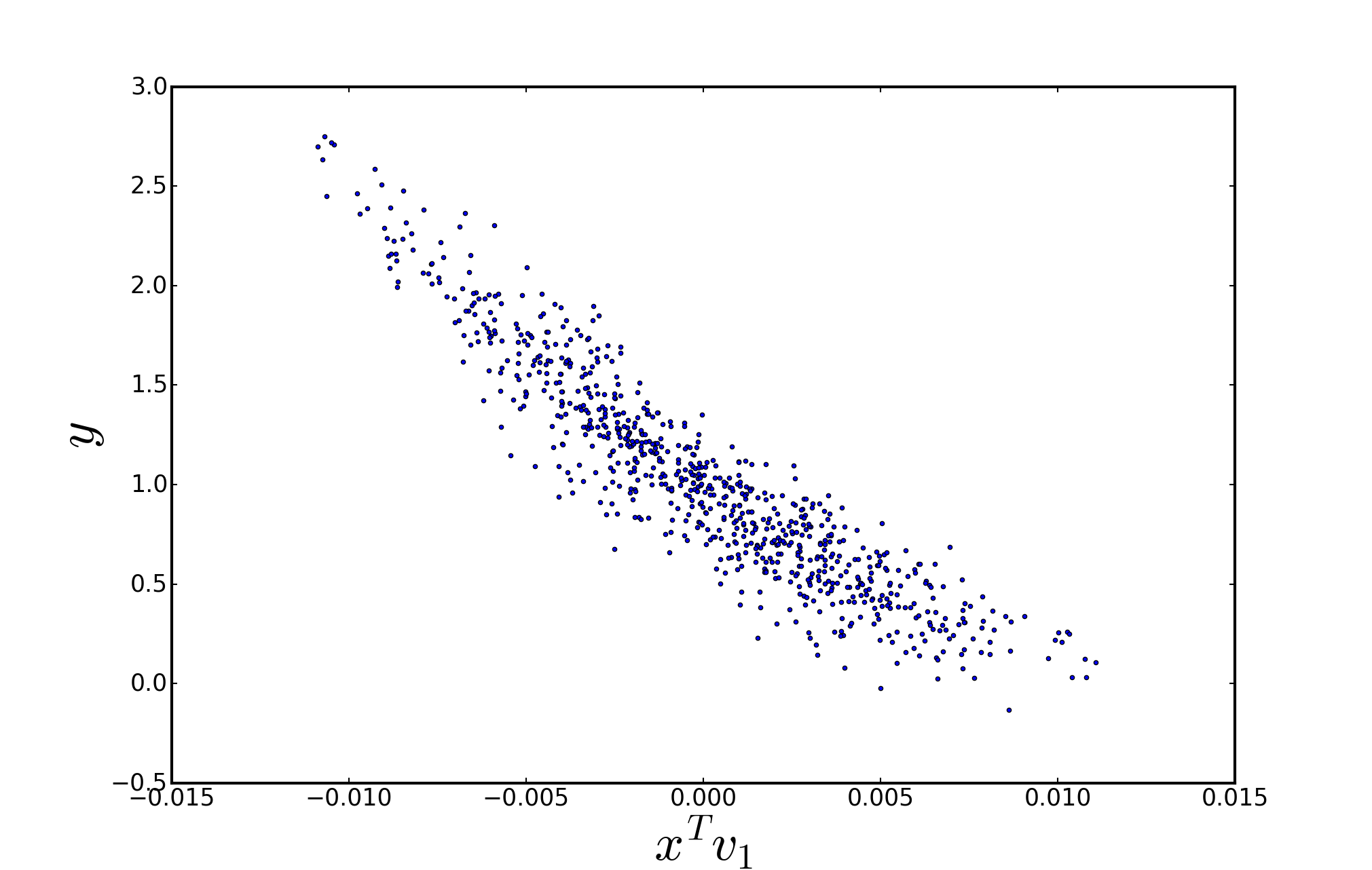}
\caption{Valeurs prises par $y$ par rapport aux r\'ealisations de $\textbf{X}$ projet\'ees sur la direction associ\'ee \`a la plus grande valeur propre donn\'ee par SIR. Malgr\'e la pr\'esence de bruit, on est en mesure d'observer un lien tr\`es clair entre les donn\'ees et la valeur de $y$}
\label{SIR_ex2}
\end{figure}

Gr\^ace \`a la figure \ref{SIR_ex2}, on peut voir que SIR, gr\^ace \`a la prise en compte des variations de $y$ lors de la recherche de directions pertinentes, est \`a m\^eme de fournir une projection qui montre une forte corr\'elation avec les donn\'ees.
L'application de SIR nous permet d'obtenir un sous-espace, form\'e par la ou les directions les plus pertinentes.
\vspace{-.4cm}
\section{D\'{e}termination de param\`{e}tres gr\^{a}ce \`{a} SIR}
\vspace{-.2cm}
Une fois que les donn\'ees son projet\'ees il faut donc lier l'espace des donn\'ees avec celui des param\`etres. Le nombre optimal de directions doit donc \^etre d\'et\'ermin\'e. Pour ce faire, nous pouvons proc\'eder visuellement ou par un protocole de validation crois\'ee, mais seulement dans l'hypoth\`ese o\`u les directions les plus pertinentes sont toujours les premi\`eres, et o\`u la dimension optimale du sous-espace est toujours la m\^eme. Une fois tous les \'echantillons de la base projet\'es, l'\'echantillon dont on souhaite conna\^itre la valeur du param\`etre est \`a son tour projet\'e. L'estimateur de la valeur du param\`etre est obtenue \`a partir de ces plus proches voisins dans le sous-espace, par une moyenne des valeurs du param\`etre correspondant \`a ceux-ci. Dans cette partie nous essaierons de retrouver gr\^ace \`a SIR les informations sur la profondeur et la largeur de raies spectrales gaussiennes bruit\'ees :

\begin{equation}\label{eq_gauss}
x(\lambda;h,w) = 1 - h \times e^{-\frac{1}{2}\left(\frac{\lambda-\lambda_0}{w}\right)^2} + \epsilon
\end{equation}
$\lambda_0=0$ est commune \`a tous les signaux. $h$ et $w$ sont uniform\'ement distribu\'es $h \sim \mathcal{U}[0.1,0.9]$, $w \sim \mathcal{U}[1,3]$ et le bruit $\epsilon$ suit une distribution gaussienne $\epsilon \sim \mathcal{N}(0,\sigma_\epsilon)$. 

Dans ce cas, une PCA (sous-optimale) sera appliqu\'ee comme un pr\'e-traitement \`a SIR de sorte \`a d\'ecro\^itre suffisamment le conditionnement de $\boldsymbol{\Sigma}$ pour permettre son inversion. La variance du bruit suppos\'ee ind\'ependante de la puissance du signal sur chaque composante de l'espace, est valable pour l'application qui nous concerne : des spectres haute r\'esolution \`a haut rapport signal sur bruit.

\subsection{D\'etermination du nombre optimal de voisins \`a consid\'erer}

Pour l'estimation de la valeur du param\`etre, une moyenne au sens des $k$ plus proches voisins a plus de sens qu'une r\'egression lin\'eaire, car pour ce qui concerne  notre probl\`eme, le lien entre donn\'ees et param\`etres n'a pas de raison d'\^etre lin\'eaire. La consid\'eration des plus proches voisins permet de palier le probl\`eme des non-lin\'earit\'es sous l'hypoth\`ese que localement les valeurs du param\`etre prises par les individus sont proches. Le choix du nombre de voisins d\'epend de la structure et de la richesse de la base de donn\'ees de r\'ef\'erence. Si l'on prend un trop grand nombre de voisins, on risque de ne plus pouvoir verifier l'hypoth\`ese de faible variabilit\'e locale ; au contraire si l'on ne consid\`ere pas suffisamment de voisins, l'estimation sera tr\`es sensible au bruit. Dans le cadre d'une base de donn\'ees \'echantillonn\'ee de mani\`ere homog\`ene dans les param\`etres, il est possible de d\'eterminer un nombre optimal de voisins quelle que soit la position dans la base (pour avoir la meilleure pr\'ecision possible compte tenu des non-lin\'earit\'es). Pour \'evaluer ce nombre optimal de voisins, nous avons proc\'ed\'e par validation crois\'ee. 


\vspace{-.4cm}

\subsection{S\'{e}lection des directions pertinentes}
\begin{figure}[H]
\centering
\includegraphics[height = 4.5cm,width =8cm]{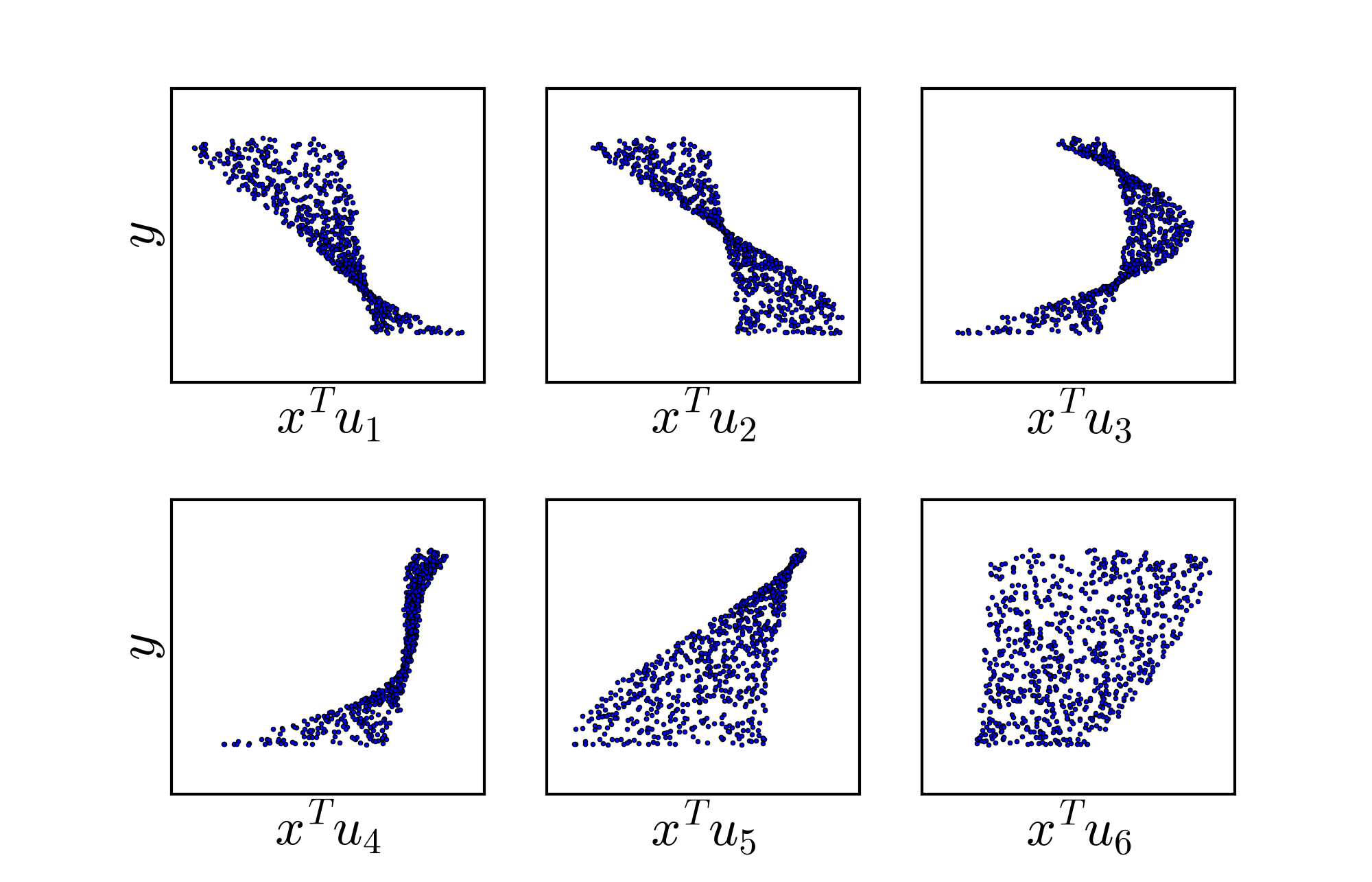}
\caption{Valeurs des param\`etres de largeur des gaussiennes $y$ par rapport aux valeurs des signaux $\textbf{X}$ projet\'es sur les 6 premi\`eres directions donn\'ees par SIR. On peut observer sur ces diff\'erents coefficients de projection, l'expression du lien non-lin\'eaire entre les \'echantillons de la gaussienne et sa largeur.}
\label{SIR_lrg_dir}
\end{figure}

\vspace{-.2cm}
Nous \'emettons l'hypoth\`ese que suivant la position de l'\'echantillon projet\'e dans le sous-espace les directions pertinentes pour sa repr\'esentation au sens du param\`etre recherch\'e ne sont pas n\'ecessairement les m\^emes. La vari\'et\'e portant l'information sur les variations du param\`etre ne serait donc pas toujours port\'ee par les m\^emes directions du sous-espace. Nous proposerons une alternative au sous-espace particulier en s\'electionnant les directions jug\'ees pertinentes pour chaque \'echantillon. Les directions choisies ne seront pas n\'ecessairement celles des vecteurs propres associ\'es aux plus grandes valeurs propres de $\boldsymbol{\Sigma}^{-1}\boldsymbol{\Gamma}$, mais celles qui seront les plus informatives au sens de la parcimonie des valeurs prises par $y$ compte tenu de la projection de l'\'echantillon \`a identifier (cf. \cite{ecmsm}).  Si l'on observe les directions obtenues lorsque l'on consid\`ere les largeurs, on obtient les r\'esultats pr\'esent\'es figure \ref{SIR_lrg_dir}.
\vspace{-.4cm}
\vspace{-.2cm}
La figure \ref{SIR_lrg_dir} montre que suivant la position de la projection d'un \'echantillon, les directions ne portent pas la m\^eme qualit\'e d'information. Par exemple un \'echantillon se projetant sur les valeurs m\'edianes de la direction $u_1$ obtient gr\^ace \`a celle-ci une information tr\`es peu pr\'ecise quant \`a la valeur de $y$. Au contraire si la valeur prise par $x^Tu_1$ est proche de la limite sup\'erieure, on est en mesure, gr\^ace \`a cette seule direction, de d\'eterminer pr\'ecis\'ement la valeur de $y$. On peut donc choisir suivant la position de l'\'echantillon dans le sous-espace quelles directions porteront l'information pertinente concernant la valeur de $y$ associ\'ee.
Dans le cas du calcul des profondeurs, le lien entre le param\`etre et le signal \'etant lin\'eaire, l'information pertinente est toujours port\'ee par les m\^emes directions de SIR.
En proc\'edant de la sorte, nous pouvons quantifier l'erreur moyenne en valeur absolue commise lors de la d\'etermination des param\`etres, et comparer ce r\'esultat avec ce qu'aurait donn\'e le meilleur sous-espace obtenu par PCA.
\vspace{-0.4cm}
\begin{table}[H]
\caption{\textnormal{Erreurs moyennes en valeur absolue lors de la d\'etermination des profondeurs et des largeurs des gaussiennes avec SIR ou la PCA.}}
\centering
\begin{tabular}{|c||p{2cm}|p{2cm}|}

\hline
 & PCA & SIR \\
\hline
\hline
Profondeurs &  0.0063 & 0.0025  \\
\hline
Largeurs &  0.029 & 0.024 \\
\hline

\end{tabular}
\vspace{0.1cm}
\label{table2}
\end{table}
Les erreurs pr\'esent\'ees en table \ref{table2}, sont calcul\'ees \`a partir de la d\'etermination des valeurs de param\`etres d'un cinqui\`eme de la base de donn\'ees, le reste servant de base de r\'ef\'erence et \`a la construction du sous-espace.
Ces r\'esultats pr\'esent\'es table \ref{table2} montrent que SIR permet d'obtenir une meilleure pr\'ecision que ce que l'on peut attendre d'une PCA telle qu'utilis\'ee dans \cite{Paletou15}.
\vspace{-.4cm}
\section{Application \`a des spectres synth\'etiques}
\begin{figure}[H]
\centering
\includegraphics[height = 4cm,width =8cm]{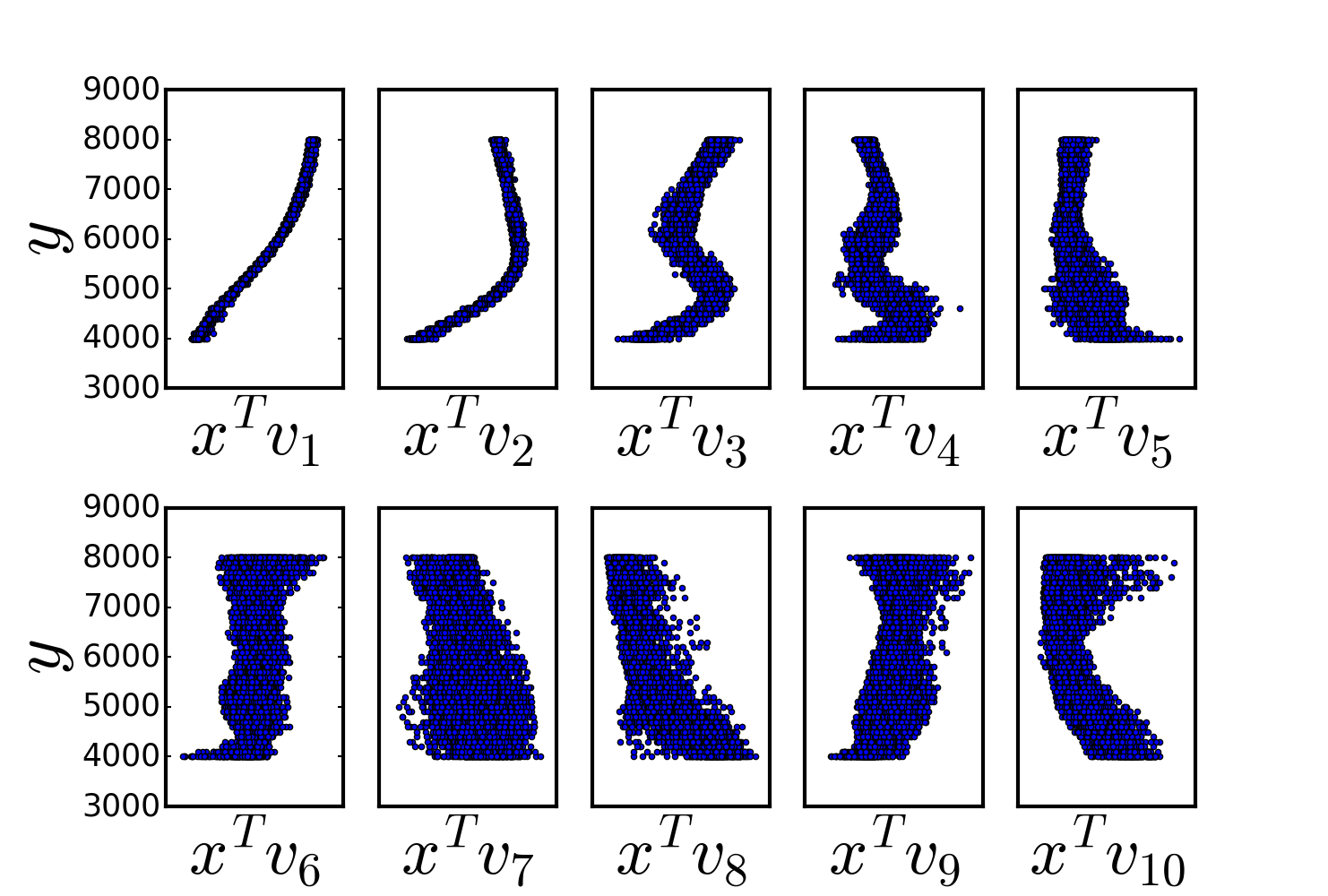}
\caption{Directions obtenues gr\^ace \`a SIR pour la d\'etermination des temp\'eratures effectives ($T_{\rm eff}$)}
\label{dir_T_sir}
\end{figure}

\vspace{-0.4cm}
Nous comparerons les r\'esultats obtenus avec ceux que l'on obtiendrait avec une PCA en suivant le protocole d\'ecrit dans \cite{Paletou15}. Comme expliqu\'e en section \ref{sec_21}, une PCA sera appliqu\'ee en pr\'e-traitement. La base de donn\'ees contient 23000 spectres synth\'etiques et chacun de ces spectres est un vecteur de 8000 points couvrant le domaine spectral $390-680~ \mathrm{nm}$ 
(cf. \cite{Paletou15}) . Le pr\'e-traitement ram\`enera cet espace \`a un espace de dimension 50 gr\^ace \`a une PCA. En termes de valeurs prises par les param\`etres (m\^emes param\`etres que dans \cite{Paletou15}), la base est \'echantillon\'ee uniform\'ement comme suit : $T_{\rm eff} \in [4000:8000]~K$, $\Delta (T_{\rm eff})=100~K$, $log(g) \in [4:5] ~ \mathrm{dex}$, $\Delta (log(g))=0.2 ~ \mathrm{dex}$, $[Fe/H] \in [-1:0.5] ~ \mathrm{dex}$, $\Delta ([Fe/H])=0.1 ~ \mathrm{dex}$, nous nous int\'eressons aussi \`a un quatri\`eme param\`etre, la vitesse de rotation projet\'ee, $v\sin (i) \in [0:100] ~ \mathrm{km/s}$, $\Delta (v\sin (i)\\=2 ~ \mathrm{km/s}$ entre 0 $\mathrm{km/s}$ et 20 $\mathrm{km/s}$ et $\Delta (v\sin (i)=5 ~ \mathrm{km/s}$ entre 20 $\mathrm{km/s}$ et 100 $\mathrm{km/s}$. ($\Delta \cdot$ repr\'esente le pas d'\'echantillonnage de la base de donn\'ees) Pour chaque calcul 1000 spectres sont al\'eatoirement extraits de la base et bruit\'es avec un bruit gaussien centr\'e et d'\'ecart-type $\sigma = 0.03$ (coh\'erent avec les donn\'ees observationnelles qui nous int\'eressent \`a terme) pour servir de base de test.
On obtient alors avec SIR pour chacun des param\`etres des directions telles que illustr\'es figure \ref{dir_T_sir}.

%

On peut observer sur la figure \ref{dir_T_sir} que les premi\`eres directions sont tr\`es informatives, de part le lien que l'on voit apparaitre entre $x^Tv_i$ et $y$,  quant \`a la valeur prise par $T_{\rm eff}$. Il est int\'eressant de noter que la direction num\'ero 5 par exemple n'apporte une information pertinente que pour le tiers sup\'erieur de $x^Tv_5$.

\begin{figure}[H]
\centering
\includegraphics[height = 4.5cm,width =8cm]{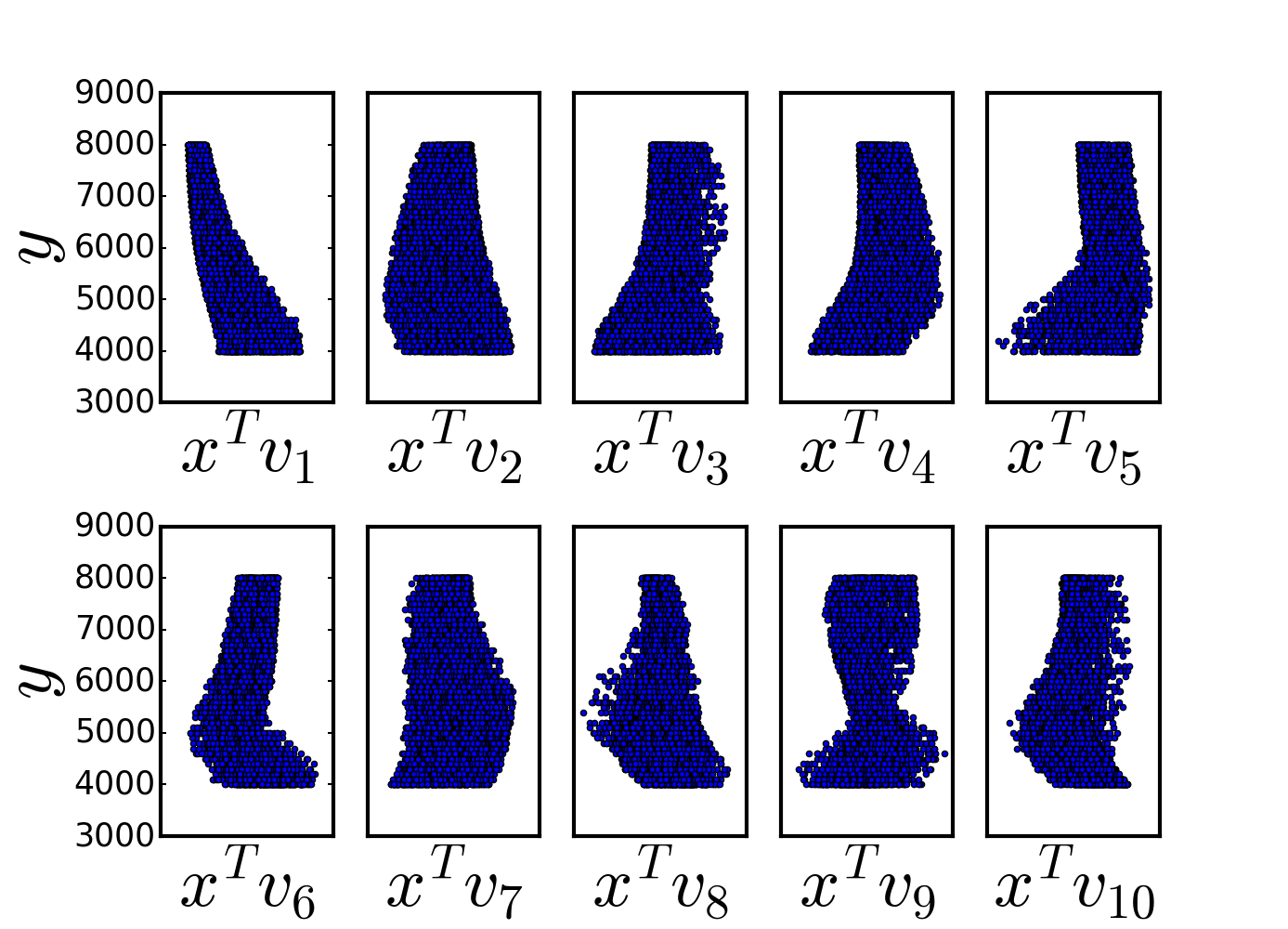}
\caption{Directions obtenues gr\^ace \`a la PCA pour la d\'etermination des temp\'eratures effectives ($T_{\rm eff}$)}
\label{dir_T_pca}
\end{figure}

On peut comparer les directions de la figure \ref{dir_T_sir} avec les projections sur les composantes obtenues avec PCA en figure \ref{dir_T_pca}.

On s'aper\c{c}oit en regardant la figure \ref{dir_T_pca}, que les directions obtenues par PCA traduisent, de part leurs construction, beaucoup moins bien la variation du param\`etre $T_{\rm eff}$ : cela signifie qu'il faut un plus grand nombre de directions pour avoir une information aussi pr\'ecise qu'avec SIR. Mais prendre plus de directions signifie aussi plus de bruit. Il est donc peut probable que la PCA seule puisse avoir des r\'esultats aussi pr\'ecis que ceux atteints gr\^ace \`a SIR.

Pour cet exemple, la base \'etant r\'eguli\`erement \'echantillonn\'ee pour tous les param\`etres, on peut d\'eterminer un nombre de voisins constant \`a consid\'erer. Une dizaine de voisins permet de rester en dessous du nombre d'\'echantillons pour une valeur d'\'echantillonage du param\`etre, nous serons ainsi peu sensibles aux non lin\'earit\'es. L'exp\'erience montre que entre 10 et 30 voisins, les r\'esultats ne sont pas diff\'erentiables


\begin{table}[H]
\caption{\textnormal{Erreurs moyennes en valeur absolue lors de la d\'etermination des param\`etres stellaires fondamentaux et du $v \sin (i)$ sur les spectres synth\'etiques.}}
\centering
\begin{tabular}{|c||p{2cm}|p{2cm}|}

\hline
 & PCA & SIR \\
\hline
\hline
$T_{\rm eff}$ &  132 K & 61 K  \\
\hline
$log(g)$ &  0.25 dex & 0.10 dex \\
\hline
$[Fe/H]$ &  0.07 dex & 0.03 dex \\
\hline
$v \sin (i)$ &  1.2 km/s & 0.09 km/s \\
\hline

\end{tabular}
\vspace{0.1cm}
\label{table1}
\end{table}

On peut observer table \ref{table1} que, comme on s'y attendait, les r\'esultats que l'on obtient avec SIR sont bien plus pr\'ecis que ceux obtenus gr\^ace \`a la PCA. L'inconv\'enient que l'on pourrait trouver \`a SIR, est que la m\'ethode dans le cas o\`u $\boldsymbol{\Sigma}$ est mal conditionn\'ee, n\'ecessite un pr\'e-traitement. Elle n\'ecessite aussi que l'on traite ind\'ependamment chaque param\`etre, ce qui rend son ex\'ecution plus lourde.
\vspace{-.4cm}
\section{Conclusion}
\vspace{-.2cm}
Nous avons vu que SIR permettait de projeter des donn\'ees sur le sous-espace qui maximise la coh\'erence de variation entre elles et un param\`etre pr\'e-d\'efini. Cette approche ne n\'ecessite pas la connaissance d'une fonction permettant de passer de l'espace des donn\'ees \`a celui du param\`etre. Il est n\'ecessaire lorsqu'on l'applique de faire attention au conditionnement de $\boldsymbol{\Sigma}$. Pour la d\'etermination de param\`etres stellaire, la m\'ethode bas\'ee sur SIR se r\'ev\`ele \^etre tr\`es efficace sur les spectres synth\'etiques par rapport \`a la m\'ethode bas\'ee sur PCA. La PCA a l'avantage, quant \`a elle, de ne n\'ecessiter qu'un espace de projection pour tous les param\`etres. Les premiers tests sur des spectres observ\'es montrent une am\'elioration d'environ 25\% sur la temp\'erature et sur la m\'etallicit\'e par rapport aux r\'esultats de PCA. Cependant les bases de spectres observ\'es \'etant souvent \'echantillonn\'ees de mani\`ere non-homog\`ene concernant certains param\`etres -certaines plages de valeurs de $log(g)$ con\-tiennent tr\`es peu d'objets par exemple-, nous sommes confront\'es \`a d'autres probl\`emes, notamment concernant la gravit\'e de surface et la vitesse de rotation projet\'ee. Un d\'eveloppement futur sera d'adapter la m\'ethode pour des bases \'echantillonn\'ees de mani\`ere non-homog\`ene.

\vspace{-0.4cm}

\bibliographystyle{unsrt}
\bibliography{IEEEabrv,References}

\end{document}